\documentstyle[aps,pre,epsfig]{revtex}

\newcommand{\be}{\begin{equation}}
\newcommand{\ee}{\end{equation}}
\newcommand{\bel}[1]{\begin{equation}\label{#1}}
\newcommand{\bea}{\begin{eqnarray}}
\newcommand{\eea}{\end{eqnarray}}
\newcommand{\ba}{\begin{array}}
\newcommand{\ea}{\end{array}}
\newcommand{\ket}[1]{\mbox{$| \, {#1}\, \rangle$}}
\newcommand{\exval}[1]{\mbox{$\langle \, {#1}\, \rangle$}}

\def\al{\alpha}
\def\de{\delta}
\def\si{\sigma}

\def\eps{\epsilon}

\def\both{\leftrightarrow}
\def\ra{\rightarrow}

\begin{document}
%\begin{large}

\title{ Integrable Markov processes  and quantum spin chains}
\author{Vladislav Popkov$^{1,2}$ and  G. M. Sch\"utz $^{1}$ }                 
\address{
$^1$ Institut f\"ur Festk\"orperforschung, Forschungszentrum J\"ulich,
52425 J\"ulich, Germany\\
e-mail: g.schuetz@fz-juelich.de
$^2$ Institute for Low Temperature Physics, 310164 Kharkov, Ukraine\\
e-mail: v.popkov@fz-juelich.de
}
\maketitle

\begin{abstract}
A set of Markov processes corresponding to systems of
 hard-core particles interacting   along the line
are shown to be solvable via a dynamic matrix product
ansatz (DMPA). We show that quantum spin Hamiltonians
can be treated by the DMPA as well, and demonstrate how
the DMPA, originally formulated for systems with open 
ends,  works for periodic systems. 
\end{abstract}

\vspace{0.5cm}
MCS numbers: 60J25, 37K10, 17B69

\vspace{2cm}

Recent advances in nonequilibrium statistics have 
allowed for the clarification of many important issues such as
exact description of nonequilibrium phase transitions,
microscopic structure of shocks, dynamical scaling etc. \cite{Schu00}.
Dynamical systems of stochastically hopping hard 
core particles (the so-called exclusion processes) 
have been at the heart of this development, in particular
due to remarkable observation that one can find exactly
the probabilities of all system configurations in a stationary
state. It was subsequently noticed that for
the totally symmetric exclusion process one can as 
well solve the evolution dynamics, namely to find conditional
probabilities $P(\nu,t|\nu^0,t^0)$ of the system being at state $\nu$
at time $t$ provided that it was in state $\nu^0$ at time
$t^0$ via coordinate Bethe Ansatz \cite{Schuetz_BA_XXX}. 
The dynamical matrix product ansatz (DMPA) proposed
in \cite{StinchcombeSchuetz} has recovered
the results of  \cite{Schuetz_BA_XXX} in a more straighforward manner. But 
for technical reasons to be explained later, application
of the DMPA was incomplete in periodic
systems.

In this communication, we show how  the DMPA is applied for 
periodic systems. Alongside, we demonstrate that the DMPA can
be used for testing the stochastic processes and/or 
one-dimensional 
spin chains for integrability. The integrability restrictions
come from the consistency conditions on the algebra generated by
the DMPA, which has  to satisfy the Yang-Baxter equation. The 
important difference from the conventional approach is that
we start directly from the Hamiltonian and do not need 
a parameter-dependent transfer matrix, commuting with it.

The continuous time 
dynamics of Markov process is defined via the master 
equation for the probability $P(\nu,t)$ of the system being at state $\nu$
at time $t$,

\bel{MasterEq}
\partial P(\nu,t) / \partial t = 
\sum_{\nu'} P(\nu',t) w_{\nu'\ra \nu} -
 P(\nu,t) \sum_{\nu'} w_{\nu \ra \nu'}
\ee

where $w_{\nu \ra \nu'}$ is the rate with which 
the system can change its configuration from 
$\nu$ to  $\nu'$.

For definiteness we consider a system of identical particles 
 on a lattice  of $L$ sites. 
Each site $k$ can be either occupied by a particle
(local occupation number $n_k=1$) or be empty
( $n_k^0=1$). The set of occupation numbers for all sites
constitutes the configuration of the system.
The system can change its configuration via  hopping of any particle to an
empty nearest neighbour site.
At  any infinitesimally small
moment of time $dt$ at most one such process can happen, 
namely the particle can hop to the empty site to the right/to the left
with the rates $D_R$ and $D_L$ respectively.
Now, consider a vector space ${\cal W} = \prod^L \otimes C^2$.  Notice that 
one can establish one-to-one correspondence between configurational space of
 our particle system and canonical basis of   ${\cal W}$; 
  $| \eta \rangle \subset {\cal W}$, with the scalar product
 $\langle \eta'| \eta\rangle = \delta_{\eta \eta'}$. 
The local particle number operators  
have form $n_k =I\otimes \ldots I\otimes e_{22}\otimes\ldots I$ , 
and $e_{11}$  respectively for holes. $ e_{ij} $ denotes a matrix 
with  $ (e_{ij})_{kl}= \de_{ik}\de_{jl}$, and it stays in the $k$-th place in 
the tensor product expression above. $I$ is the unit $2 \times 2$ matrix.  

With the help of above notations the set of equations (\ref{MasterEq})
can be rewritten as single vector equation ( we ask the  reader to refer e.g. 
to \cite{Schu00} for the details of the scheme).
 
 \bel{MasterVector}
{d |P (t) \rangle \over d t }= - H  |P (t)\rangle 
\ee
where 
$| P(t) \rangle  = \sum_{\eta \subset {\cal W}} P_\eta(t) \  | \eta \rangle $,
$H_{\eta \eta'}= -w_{\eta'\rightarrow \eta}  ,\ \ 
 H_{\eta \eta}= \sum_{\eta'} w_{\eta \rightarrow \eta'} $.

Due to the fact that the hopping is restricted to the nearest-neighbouring sites,
 the Hamiltonian above is the sum of local two-site Hamiltonians,
$H= \sum_{n=1}^L h_{n,n+1}$
 given by
\bel{h_stochastic}
4 h_{n,n+1}= 
D_R \left(
(1-\si_n^z)(1+\si_{n+1}^z)- \si_n^+ \si_{n+1}^-
\right)+ 
 D_L \left(
(1+\si_n^z)(1-\si_{n+1}^z)- \si_n^- \si_{n+1}^+
\right) 
\ee
where $\si^z, \ \si^\pm= \si^x \pm i \si^y$ are Pauli matrices.

Note that (\ref{MasterVector}) has a form of a Schr\"odinger equation 
in imaginary time, and the Hamiltonian (\ref{h_stochastic})
the form of a  nonhermitian spin Hamiltonian with nearest neighbour exchange
 interaction.

 The global Hamiltonian $H$ possesses an U(1) 
symmetry, $[H,\sum_n \si_n^z]=0$. 
The general hermitian spin Hamiltonian satisfying the latter 
relation is $XXZ$ Hamiltonian,
\bel{XXZ}
H= J  \sum_n \si_n^x \si_{n+1}^x +  \si_n^y \si_{n+1}^y 
+ \Delta  \left( \si_n^z \si_{n+1}^z - I \right). 
\ee

Thus Schr\"odinger quantum problem and Markov process for interacting
particles are closely connected. In the following we show how to
find the spectrum of $H$ and how to single
out integrable models.

% \section{Dynamic Matrix Product Anzatz (DMPA) for periodic systems}

To demonstrate how the DMPA works, we start from the Hamiltonian of the 
form
\bel{both}
H=\sum_{n=1}^L h_{n,n+1}; \ \ h= 
\left(
\begin{array}{cccc} 
0 & 0& 0 & 0\\
0 & f & u  & 0\\
0 & v& g & 0\\
0 & 0& 0 & 0
\end{array}
\right)
 \ee
with arbitrary entries $f,g,u,v$, which contains both the stochastic  
(\ref{h_stochastic}) and quantum spin system
(\ref{XXZ}). We shall consider a system  on a ring with $L$ sites.
%Along the lines of \cite{StinchcombeSchuetz},
One proposes to search for time-dependent probability 
distribution vector $|P(t)\rangle $ in the following form:

\bel{MPS}
|P (t) \rangle \sim \mbox{Tr }
\left[
\left(\ba{c} E \\ D \ea \right)^{\otimes L-1}
\otimes \left(\ba{c} EQ \\ D Q  \ea  \right)
\right]
\ee
where $E,D, Q$ are some time-dependent matrices.
The probability of a given configuration $\nu$ is given by the scalar product 
of the above vector with the basis vector $|\nu\rangle$ of the configuration,
$P(\nu,t)= \langle \nu| P(t) \rangle$. For example, 
probability $P_{\eta}(t)$ of a state $\eta = \{ 0\ 0\ A\ 0\ A \} $ for a system 
of 5 sites is sought 
in the form $P_{\eta}(t) \sim \mbox{Tr } \left( E\ E\ D\ E\ D\  \ Q \right)$,
up to normalization.

One  requires matrices $E,D, Q$ such that 
(\ref{MPS}) satisfies the master equation  (\ref{MasterVector}).
To simplify matters, first take $Q$ in (\ref{MPS}) to be a unit matrix $Q=I$ 
(this would allow to retrieve the translationally invariant part of the 
$H$ spectrum as we shall see below). 
Given that, and that the 
 Hamiltonian in (\ref{MasterVector}) splits into local ones,
 $H= h_{1,2}+h_{2,3}+ \ldots h_{L-1,L} + h_{L,1}$,
it is sufficient for the purpose that the matrices $E,D$ satisfy
quadratic relation

\bel{DMPA_4}
\left(
\frac{1}{2} \frac{d}{dt} + {\bf h} 
\right)
{E \choose D } \otimes {E \choose D } = 
{X^0 \choose X } \otimes{E \choose D } -  {E \choose D } \otimes {X^0 \choose X }
\ee
conveniently written in short form 
 \bel{short}
\left( \frac{1}{2}\frac{d}{dt} + h \right) \ket{\bf A} \otimes \ket{\bf A}
= \ket{\bf X}\otimes \ket{\bf A} -  \ket{\bf A} \otimes \ket{\bf X}.
\ee
where $\langle {\bf A} | = ( E \ D )$, 
 $\langle {\bf X} | = ( X^0 \ X )$.

For $Q \neq I$ which is generic case
 one gets two additional relations in addition to (\ref{short}),
connected with the action of the terms $ h_{L-1,L}, h_{L,1}$
of the Hamiltonian:
 \bea
\label{I}
\left( \frac{1}{2}\frac{d}{dt} + h \right) \ket{\bf A} \otimes \ket{{\bf A} Q} &=& 
 \ket{\bf X}\otimes \ket{{\bf A} Q} -  \ket{\bf A} \otimes \ket{{\bf X}Q}\\
\label{II}
\left( \frac{1}{2}\frac{d}{dt} + h \right) \ket{{\bf A}Q} \otimes \ket{\bf A} &=& 
 \ket{{\bf X}Q}\otimes \ket{{\bf A}} -  \ket{{\bf A}Q} \otimes \ket{{\bf X}}
\eea

Equation (\ref{short}) gives the relations for the matrices 
$E,D$ while  Eqs. (\ref{I},\ref{II}) impose 
restrictions on $Q$.

 Equation (\ref{DMPA_4}) has four components:
\bea
\label{eq1}
\frac{1}{2}\left(\dot{E}E + E\dot{E}\right)  &=&  X^0E - EX^0 \\
\label{eq2}
\frac{1}{2}\left(\dot{E}D + E\dot{D}\right) + f  E D  + u DE 
                                            & =&  X^0D - EX\\
\label{eq3}
\frac{1}{2}\left(\dot{D}E + D\dot{E}\right)  + g D E + vED
                                            & =&  X E - D X^0 \\
\label{eq4}
\frac{1}{2}\left(\dot{D}D + D\dot{D}\right)  &=&  XD - DX 
\eea
Using the reduction 
\bel{Edot}
\dot{E}=0;\ X^0=0,
\ee
Eq. (\ref{eq1}) is satisfied trivially.
Supposing additionally that  ${E}^{-1}$ exists, 
we obtain from  (\ref{eq2}), (\ref{eq3}): 
\bel{X}
2 X =  (g-f) D + vED E^{-1} -u E^{-1}D E;
\ee
\bel{Ddot}
 \dot{D} + (g+f) D + vED E^{-1} + u E^{-1}D E=0
\ee

Substituting $X,\dot{D}$ into  (\ref{eq4}), one gets:
\bel{DD}
(g+f) DD + u D E^{-1}D E  + vE D E^{-1} D =0 
\ee

Now, consider the  two equations (\ref{I},\ref{II}) with 
$Q$ present.
First, suppose that $Q$ is time-independent, and $Q^{-1}$ exists.
Then, Eq.(\ref{I}) is satisfied with the choices (\ref{X}-\ref{DD})
since multiplication of  (\ref{I}) by $Q^{-1}$ from the right reduces 
(\ref{I}) to (\ref{short}). In order to satisfy Eq.(\ref{II}) with
the choice (\ref{Edot}-\ref{DD}),the matrix $Q$ must obey
\bel{Q}
[Q,DE^{-1}] = [Q,E^{-1}D] = 0.
\ee
Note that in general we assume that  
$Q$ does not commute neither with $D$ nor with $E$ separately, 
$[Q,E]\neq 0; \ \  [Q,D]\neq 0$.

Proceeding along the lines \cite{StinchcombeSchuetz}, we can get rid of 
time-dependence of $D$ via a formal Fourier Transform
\bel{FT}
{\cal D}_p(t)  =  \sum_k  \mbox{e}^{ipk} E^{k-1}D(t) E^{-k} 
\ee
Indeed, it follows from (\ref{FT}) that 

\bel{EDE}
E^{-1} {\cal D}_p(t) E =  \mbox{e}^{ip} {\cal D}_p(t); \ 
\ee

Substituting  (\ref{FT}), (\ref{EDE}) into (\ref{Ddot}), one gets

$$ \dot{\cal D}_p + \eps_p {\cal D}_p= 0; \ {\cal D}_p(t)=e^{-\eps_p t}{\cal D}_p(0)
$$ 
with the ``dispersion relation''
\bel{dispersion}
 \eps_p= g+f + v \mbox{e}^{-ip} + u \mbox{e}^{ip}
\ee

The inverse Fourier Transform is 
\bel{InvFT}
D(t) = E \frac{1}{2\pi} \int_{-\pi}^{\pi} \mbox{d}p {\cal D}_p(t)=
 E \frac{1}{2\pi} \int_{-\pi}^{\pi} \mbox{d}p  \mbox{e}^{-\eps_p t} {\cal D}_p(0)
\ee
In what follows we shall omit the time argument 
in initial matrices   ${\cal D}_p(0)$.
Inserting (\ref{InvFT}) into (\ref{eq4}), and using (\ref{EDE}),
one obtains:

\bel{FTeq4}
 {E^2  \over 2 \pi} \int_{-\pi}^{\pi} \int_{-\pi}^{\pi}
\mbox{e}^{-(\eps_{p_1}+\eps_{p_2} )t}
{\cal D}_{p_1}{\cal D}_{p_2} a(p_1,p_2) \mbox{d}p_1 \mbox{d}p_2 = 0
\ee
where
\bel{a}
 a(p_1,p_2) = u \mbox{e}^{i p_1 + i p_2} + (f+g) \mbox{e}^{ i p_1}  + v
\ee

Now, integral in  (\ref{FTeq4}) may be split in two parts as 
\bel{split}
 \int_{-\pi}^{\pi} \int_{-\pi}^{\pi} \ldots \mbox{d}p_1 \mbox{d}p_2=
 \int_{-\pi}^{\pi} \int_{-\pi}^{p_1} \ldots +
\int_{-\pi}^{\pi} \int_{p_1}^{\pi} \ldots= I_{p2<p1}+ I_{p2>p1}  
\ee
By changing the order of integration and interchanging $p_1 \both p_2$
in the last term in (\ref{split}) we obtain from (\ref{FTeq4}):

$$
\int_{-\pi}^{\pi} \int_{-\pi}^{p_1} \left( 
a(p_1,p_2){\cal D}_{p_1} {\cal D}_{p_2}+ a(p_2,p_1){\cal D}_{p_2}{\cal D}_{p_1}
\right) \mbox{e}^{-(\eps_{p_1}+\eps_{p_2} )t}
 \mbox{d}p_1 \mbox{d}p_2 =0
$$

Since this is satisfied at all times $t$, we must require the expression
inside the integrand to vanish
\bel{Dcomm}
{\cal D}_{p_1} {\cal D}_{p_2} = -   {a(p_2,p_1) \over a(p_1,p_2)}
 {\cal D}_{p_2}{\cal D}_{p_1}
\ee

Finally, we require the  matrix $Q$ from (\ref{I},\ref{II})
to satisfy
$[Q,E^{k-1} D E^{-k}]=0 $, which contains (\ref{Q}) as a special case and yields
\bel{QDp}
[Q,{\cal D}_p (t)]=0 
\ee

%\section{Calculation of expectation values and Bethe Ansatz equations.}
To obtain
 Bethe Ansatz equations for the spectrum,
consider the quantity 
\bel{DDDD}
\mbox{Tr } ({\cal D}_{p_1} {\cal D}_{p_2}\dots   {\cal D}_{p_N} E^{L} Q),
\ee
which is the building block of the expression for expectation values. Indeed,
for a system with just two particles 
 the  expectation value $\exval{n_x(t)n_y(t)}$, according to (\ref{MPS}),
(\ref{EDE}),(\ref{InvFT}),
can be written as  
\bea
\nonumber
\exval{n_x(t)n_y(t)} = \mbox{Tr } (E^{x-1} D E^{y-x-1} D 
E^{L-y} Q)/Z_L=& & \\
\nonumber
\int \mbox{d}p_1\int d\mbox{p}_2
\mbox{e}^{-(\epsilon_{p_1}+\epsilon_{p_2})t} \mbox{e}^{-ip_1x-ip_2y}
\mbox{Tr } ({\cal D}_{p_1}{\cal D}_{p_2}E^L Q)/Z_L, & &
\eea
where $Z_L$ is an appropriate normalization factor,
$
Z_L = \sum_{x<y =1,2,\ldots L} \exval{n_x(t)n_y(t)}
$.
 Generally,
one is interested in a system with $N$ particles which leads to 
(\ref{DDDD}).

Using  (\ref{Dcomm}), and commuting ${\cal D}_{p_k}$ around the circle, one obtains using 
(\ref{EDE}), (\ref{QDp}) and the cyclic invariance  property of the trace:
$$
\mbox{Tr } ({\cal D}_{p_1} {\cal D}_{p_2}\dots   {\cal D}_{p_N} E^{L} Q)=
(-1)^{N+1} \prod_{n=1}^N \frac { a(p_n,p_k)} { a(p_k,p_n)} e^{ip_k L}
\mbox{Tr } ({\cal D}_{p_1} {\cal D}_{p_2}\dots   {\cal D}_{p_N} E^{L} Q)
$$
and consequently
\bel{BA4}
(-1)^{N+1} \prod_{n=1}^N \frac { a(p_n,p_k)} { a(p_k,p_n)} \mbox{e}^{ip_k L}=1
\ee

One recognizes in (\ref{BA4}), (\ref{dispersion})
the well-known expressions for the spectrum of 
the $XXZ$ spin model, for  
$f=g=-2J\Delta$, $u=v= 2 J$.   

It may seem that the proposed line of argument is not changed if 
the auxiliary matrix $Q$ that we have introduced  is chosen
to be a unity matrix $Q=I$. This  trivial choice however would 
bring about an additional 
relation. Namely let us rewrite (\ref{DDDD})  as

\bel{Q=I}
\mbox{Tr } ({\cal D}_{p_1} {\cal D}_{p_2}\dots   {\cal D}_{p_N} E^{L})=
\mbox{Tr } ({\cal D}_{p_1} {\cal D}_{p_2}\dots   {\cal D}_{p_N} E  E^{L-1})
\ee
and then commute single  matrix E around the circle. We shall obtain
\bel{impulse_zero}
\mbox{e}^{ip_1+ip_2+\ldots ip_N}= 1
\ee
then. This  means putting $Q=I$ (or more generally allowing for
$[Q,E]=0 $) we recover  only the part of spectrum
 with the total  quasimomentum  equal to zero, $p_1+p_2+\ldots p_N=0$.  

We have thus shown how from the Dynamis Matrix Product Ansatz one obtains
the spectrum for spin $1/2$ $XXZ$ chain. Note that until now 
we have not got any restriction on parameters in (\ref{both}).
 This is a consequence of the fact that  
the Hamiltonian (\ref{both}) is integrable for any choice of parameters.
In the following we show how the
the  integrability restrictions enter in a more general setting. 

%\section{Systems of spin $1$ and Yang-Baxter integrability conditions.}
Consider nearest neighbour quantum spin 1 Hamiltonians. 
This case corresponding to three states at a  site, e.g. $n_i=0,1,2$
comprises  many integrable systems (see e.g. a list given in \cite{19v} where however
several symmetry restrictions are imposed).  With those restrictions being relaxed, 
there are many more, e.g. \cite{other_models,Ibragim}.

To be specific, we shall consider the hermitian Hamiltonian of the 
form 

\bel{h_spin1}
h = \left( \ba{ccccccccc}
0 &   0   &   0   &   0   & 0 &   0   &   0   &   0   & 0 \\
0 & a&   0   &g_{A}& 0 &   0   &   0   &   0   & 0 \\
0 &   0   & b&   0   & 0 &   0   &g_{B}&   0   & 0 \\
0 &g_{A}&   0   & a& 0 &   0   &   0   &   0   & 0 \\
0 &   0   &   0   &   0   & d &   0   &   0   &   0   & 0 \\
0 &   0   &   0   &   0   & 0 & c&   0   &g& 0 \\
0 &   0   &g_{B}&   0   & 0 &   0   & b&   0   & 0 \\
0 &   0   &   0   &   0   & 0 &g&   0   & c& 0 \\
0 &   0   &   0   &   0   & 0 &   0   &   0   &   0   & e \ea \right)
\ee

with $8$ independent parameters. If $g=g_{B}=g_{A}=1$,
$d=e=0$ and $a=b=c=-1$, it reduces to 
isotropic Hamiltonian of Sutherland \cite{Sutherland} $h=P-I$, where $P$ is permutation operator 
in $SU(3)\otimes SU(3)$ defined by $P (A \otimes B ) =  (B\otimes A ) P $ for
any matrices $A,B$ from $SU(3)$.
Again, the complete Hamiltonian has the form $H=\sum_{n=1}^L h_{n,n+1} $ and
we are interested in a solution of Schr\"odinger equation 
$i {d \over dt}| P\rangle = H |P\rangle $. 

Analogously to (\ref{MPS}) one proposes an Ansatz for $|P (t) \rangle $ as

\bel{MPS_spin1}
|P (t) \rangle \sim \mbox{Tr }
\left[
\left(\ba{c} E \\ D^A \\ D^B \ea \right)^{\otimes L-1}
\otimes \left(\ba{c} EQ \\ D^A Q \\ D^B Q \ea  \right)
\right]
\ee
with  matrices  $D^A, D^B$ referring to local  variable $n_i=0,1,2$ respectively.

Ansatz  (\ref{MPS_spin1}) leads to the equations (\ref{short},\ref{I},\ref{II}) 
where now  $\langle {\bf A} | = ( E \ D^A \ D^B )$, 
 $\langle {\bf X} | = ( X^0 \ X^A \ X^B )$. 
Eq.(\ref{short}) has 9 components,
written analogously to (\ref{eq1}-\ref{eq4}). The first one is satisfied
with the choice (\ref{Edot}).
Those four that are
linear in $E$, yield analogically to (\ref{X},\ref{Ddot}):
 \bel{A,X}
 \dot{D}^A+ 2 a {D}^A + 
g_{A}\left( 
 E^{-1}D^A E +  ED^A E^{-1}
\right)
\ee
$$ 
2  X^A =  g_{A} \left( ED^A E^{-1} - E^{-1}D^A E \right);
$$
and the Eq. for $ \dot{D}^B,  X^B$ is obtained by exchanging $a\rightarrow b,
\ A \rightarrow B$. The additional Eqs.(\ref{I},\ref{II}) lead to the conditions 
$[Q,E^{k-1} D^Z(t) E^{-k}]=0$,  $k=0,\pm 1, \ldots, Z=A,B$  
on $Q$-matrix.

Eliminating time  derivatives in the four remaining equations 
that are quadratic in $D^Z$ we obtain:
\bea
\label{DD1}
(2a-d) D^A D^A + g_{A} \left(  D^A E^{-1}D^A E +  E D^A E^{-1} D^A
        \right) = 0 & &\\
\label{DD2}
(a+b-c) D^A D^B - g  D^B D^A + g_{A} E D^A E^{-1} D^B + g_{B}  D^A E^{-1}D^B E
 = 0 & &\\
\label{DD3}
(a+b-c) D^B D^A - g  D^A D^B + g_{A} E D^B E^{-1} D^A + g_{A}  D^B E^{-1}D^A E
 = 0 & &\\
\label{DD4}
(2b-e) D^B D^B + g_{B} \left(  D^B E^{-1}D^B E +  E D^B E^{-1} D^B
        \right) = 0 & &
\eea

One proceeds analogically then to what we have done in the spin $1/2$ case.
Introducing generalized Fourier Transform
${\cal D}^A_p(t)  =  \sum_k \al^k \mbox{e}^{ipk} E^{k-1}D^A(t) E^{-k} $,
${\cal D}^B_p(t)  =  \sum_k \beta^k \mbox{e}^{ipk} E^{k-1}D^B(t) E^{-k} $
leads to eliminating the time-dependence in (\ref{A,X}):
$ \dot{\cal D}^Z_p + \eps_p {\cal D}^Z_p= 0; \ {\cal D}^Z_p(t)=e^{-\eps^Z_p t}
{\cal D}^Z_p(0); \ Z=A,B,$ where 
\bel{eps_A_B}
\eps^A_p= 2 a + g_A (\al \mbox{e}^{ip} +  \mbox{e}^{-ip}/\al),\ 
\eps^B_p= 2 b + g_B (\beta \mbox{e}^{ip} +  \mbox{e}^{-ip}/\beta).
\ee
The inverse Fourier Transform 
$D^{A,B}(t) = E \frac{1}{2\pi} \int_{-\pi}^{\pi} \mbox{d}p {\cal D}^{A,B}_p(t)$,
being inserted into (\ref{DD1}-\ref{DD4}), yields 

\bea
\label{algebraAA}
0 & = & \int \mbox{d}p_1 \int \mbox{d}p_2 a_{12} {\cal D}^A_{p_1}
{\cal D}^A_{p_2} \mbox{e}^{-(\epsilon^A_{p_1}+\epsilon^A_{p_2})t}\\
\label{algebraAB}
0 & = & \int \mbox{d}p_1 \int \mbox{d}p_2 [c_{12} {\cal D}^A_{p_1}
{\cal D}^B_{p_2} - g \beta \mbox{e}^{ip_2} 
{\cal D}^B_{p_2}{\cal D}^A_{p_1}]
\mbox{e}^{-(\epsilon^A_{p_1}+\epsilon^B_{p_2})t}\\
\label{algebraBA}
0 & = & \int \mbox{d}p_1 \int \mbox{d}p_2 [d_{12} {\cal D}^B_{p_1}
{\cal D}^A_{p_2} - g \al \mbox{e}^{ip_2} 
{\cal D}^A_{p_2}{\cal D}^B_{p_1}]
\mbox{e}^{-(\epsilon^B_{p_1}+\epsilon^A_{p_2})t}\\
\label{algebraBB}
0 & = & \int \mbox{d}p_1 \int \mbox{d}p_2 b_{12} {\cal D}^B_{p_1}
{\cal D}^B_{p_2} \mbox{e}^{-(\epsilon^B_{p_1}+\epsilon^B_{p_2})t}
\eea

with the functions
\bea
\label{a12}
a_{12} \equiv a(p_1,p_2) & = & 
\al \mbox{e}^{ip_1} (2 a -d) +  g_{A} (1+\al^2 \mbox{e}^{ip_1+ip_2}) \\
\label{b12}
b_{12} \equiv b(p_1,p_2) & = & 
\beta \mbox{e}^{ip_1} (2 b-e) +  g_{B} (1+\beta^2 \mbox{e}^{ip_1+ip_2}) \\
\label{c12}
c_{12} \equiv c(p_1,p_2) & = &  (a+b-c) \al  \mbox{e}^{ip_1} +
 g_{B} \al \beta \mbox{e}^{ip_1+ip_2} +  g_{A} \\
\label{d12}
d_{12} \equiv d(p_1,p_2) & = &  (a+b-c) \beta  \mbox{e}^{ip_1} +
 g_{A} \al \beta \mbox{e}^{ip_1+ip_2} +  g_{B} 
\eea

The trick with the $p_1$ and $p_2$ exchange  in 
(\ref{algebraAA}-\ref{algebraBB}) leads consequently (see  (\ref{Dcomm})
and the paragraph above it) to

\bea
\label{DA_DA}
 {\cal D}^A_{p_1} {\cal D}^A_{p_2}=-{ a_{21} \over  a_{12} } 
{\cal D}^A_{p_2}{\cal D}^A_{p_1};\ \ & &{\cal D}^B_{p_1}
 {\cal D}^B_{p_2}= - {b_{21}\over b_{12}}
{\cal D}^B_{p_2}{\cal D}^B_{p_1}
 \\
\label{DA_DB_0}
c_{12} {\cal D}^A_{p_1}{\cal D}^B_{p_2}
 - g \beta \mbox{e}^{ip_2}{\cal D}^B_{p_2}{\cal D}^A_{p_1}  &+& 
c_{21} {\cal D}^A_{p_2}{\cal D}^B_{p_2} 
- g \beta \mbox{e}^{ip_1}{\cal D}^B_{p_1}{\cal D}^A_{p_2} =0  \\
\label{DB_DA_0}
d_{12} {\cal D}^B_{p_1} {\cal D}^A_{p_2} 
- g \al \mbox{e}^{ip_2} {\cal D}^A_{p_2}{\cal D}^B_{p_1} &+&
d_{21} {\cal D}^B_{p_2} {\cal D}^A_{p_1} 
- g \al \mbox{e}^{ip_1} {\cal D}^A_{p_1}{\cal D}^B_{p_2} =0  
\eea
and a restriction $g_B=\al/\beta g_A;\  \al=\pm \beta$. The latter 
one is a consequence of the constraint 
$\eps^A_{p_1} + \eps^B_{p_2} = \eps^B_{p_1} + \eps^A_{p_2}$, 
needed to equate the time-dependent factors after $p_1 \both p_2$ exchange 
in   (\ref{algebraAB},\ref{algebraBA}).

Eqs. (\ref{DA_DB_0},\ref{DB_DA_0}) can be written in the form 
\def\denom{c_{12} d_{12} - g^2 \al \beta  \mbox{e}^{2ip_1 } }
\bea
\label{DA_DB}
{\cal D}^A_{p_1} {\cal D}^B_{p_2} = {\cal D}^B_{p_2} {\cal D}^A_{p_1} \ \ 
{ d_{12} g \beta \mbox{e}^{ip_2} -  d_{21} g \beta \mbox{e}^{ip_1} 
\over
\denom}
+ {\cal D}^A_{p_2} {\cal D}^B_{p_1}\ \ 
{ g^2 \al \beta  \mbox{e}^{ip_1+ip_2} -  d_{12} c_{21}
\over
\denom}
 & & \\
\label{DB_DA}
{\cal D}^B_{p_1} {\cal D}^A_{p_2} = {\cal D}^A_{p_2} {\cal D}^B_{p_1} \ \ 
{ c_{12} g \al \mbox{e}^{ip_2} -  c_{21} g \al \mbox{e}^{ip_1} 
\over
\denom}
+ {\cal D}^B_{p_2} {\cal D}^A_{p_1}\ \ 
{ g^2 \al \beta  \mbox{e}^{ip_1+ip_2} -  c_{12} d_{21}
\over
\denom}
\eea
The Eqs.(\ref{DA_DA},\ref{DA_DB},\ref{DB_DA})
 between the Fourier components ${\cal D}^A_p,{\cal D}^B_p$ define an algebra 
conveniently written via a scattering matrix $S$ as
\bel{S}
{\cal D}_{p_1}^Z {\cal D}_{p_2}^{Z'} =
S_{Y Y'}^{Z Z'}(p_1,p_2)  {\cal D}_{p_2}^{Y'} {\cal D}_{p_1}^Y
\ee
Applying it twice we see that it  satisfies the inversion relation
($S^T$ means transposition) 
$S(p_1,p_2) S^T(p_2,p_1) =1$. 

Now, note that application of (\ref{S})  exchange the order of indexes $p_1$ and $p_2$.
For the cubic terms
$ {\cal D}_{p_1}^X {\cal D}_{p_2}^Y  {\cal D}_{p_3}^Z$ the  exchange 
of order of indexes $ 123 \ra 321$ can be done either by subsequent series of 
$1\both 2,\ 1\both 3, \ 2\both 3$  or
$2\both 3,\ 1\both 3, \ 1\both 2$   exchanges. Associativity  of the algebra
requires the result to be the same in both cases, which amounts to
\bel{YBE}
S^{ij}_{i''j''}(p_1,p_2) S^{i''k}_{i'k''}(p_1,p_3) S^{j''k''}_{j'k'}(p_2,p_3) =
S^{jk}_{j''k''}(p_2,p_3) S^{ik''}_{i''k'}(p_1,p_3) S^{i''j''}_{i'j'}(p_1,p_2) .
\ee
This is the Yang-Baxter equation (YBE). Summing over repeating indexes is
implied. The YBE guarantees the consistency of the 
algebra on the cubic and further levels.  

The YBE impose severe restrictions on the $S$ matrix elements, which 
depend on the coefficients of the Hamiltonian (\ref{h_spin1}).
Nontrivial solutions of the YBE  single out integrable models. 
We have investigated  (\ref{YBE})  with the help of Mathematica, and found the
following solution:

\bea
\nonumber
\al= i_1 \beta,\ \ \  
 g_B=i_1 g_A,\ \ \ & &
g_A = i_2 g,\  \ \ 
d= 2 a + (i_3 -1) g,\  \ \ 
 \\
\nonumber
%\label{sol}
e= 2 b + (i_4 -1) g,\ \ \  & &
c= a+b+ i_5 g,  
\eea

%\bel{s}
%\begin{array} {l l l}
%\al= i_1 \beta & g_B=i_1 g_A & g_A = i_2 g\\ 
%d= 2 a + (i_3 -1) g & e= 2 b + (i_4 -1) g & c= a+b+ i_5 g
%\end{array}
%\ee
where $i_1, i_2,i_3, i_4,i_5 = \pm 1$.

Sutherland's \cite{Sutherland} and, more generally,
solution $\#$ 4 of \cite{19v} is contained in the above.
To the best of our knowledge, the general manifold describes
a new solution to the YBE.

To summarise, the Dynamic Matrix Product State method turns out to be 
an effective tool 
to check the integrability of both stochastic processes and 
quantum Hamiltonians. 
Applying it saves the conventional tedious step of calculating 
parameter-dependent transfer matrices satisfyng the the Yang-Baxter equation,
commuting with the original Hamiltonian.  E.g. for investigating
spin $1$ chains conventionally, we would need first to solve YBE with 
in principle $27^2$ 
independent components. In the DMPA approach  we needed only the original Hamiltonian
to start with, and then additionally to solve at most $8^2$ equations (\ref{YBE})
for the algebra associativity. (In fact the number of equations is 
much  smaller
due to symmetries).   At the same time, we believe that for 
all the integrable models obtained 
the commuting transfer-matrices do exist, although we do not need them directly. 
Also, Bethe Ansatz equations for the spectrum can be constructed directly 
from the algebra, as was demonstrated for $XXZ$ model (see (\ref{BA4})).
For the other models  the  Bethe Ansatz equations are
readily constructed \cite{Ibragim} by  following e.g. the algebraic approach \cite{Kulish}.
   
The DMPA approach does not make reference to the actual representations of the 
matrices involved. However, it would 
be interesting to find them. In special cases (namely, when they are not
time dependent, which means only stationary (ground) state is investigated) 
the representations are known  \cite{Derrida,EssRit,Rittenberg}.

%\end{large}
\end{document}